\documentclass[twocolumn]{aastex6}
\usepackage{apjfonts}
\usepackage{graphicx,amssymb,natbib}
\usepackage{epstopdf}
\usepackage{amsmath}

\newcommand{\kms}{km~s$^{-1}$}
\newcommand{\perbeam}{beam$^{-1}$}
\newcommand{\etal}{et al.}

\newcommand{\Halpha}{H$\alpha$}
\newcommand{\HI}{\mbox{H\textsc{i}}}

\shortauthors{Hallenbeck \etal}

\begin{document}
\title{HI in Virgo's ``Red and Dead'' Dwarf Ellipticals ---\\ A Tidal Tail and Central Star Formation}
\author{Gregory Hallenbeck\altaffilmark{1,2}, Rebecca Koopmann\altaffilmark{1}, Riccardo Giovanelli\altaffilmark{3}, Martha P. Haynes\altaffilmark{3}, Shan Huang\altaffilmark{4}, Lukas Leisman\altaffilmark{3}, Emmanouil Papastergis\altaffilmark{5}}

\altaffiltext{1}{Union College, Department of Physics \& Astronomy{, 807 Union Street, Schenectady NY 12308}; hallenbg@union.edu, koopmanr@union.edu}
\altaffiltext{2}{Washington \& Jefferson College, Department of Computing and Information Studies, {60 S Lincoln Street, Washington PA, 15301}.}
\altaffiltext{3}{Cornell Center for Astrophysics and Planetary Science (CCAPS), Space Sciences Building, Cornell University, Ithaca, NY 14853; riccardo@astro.cornell.edu, haynes@astro.cornell.edu, leisman@astro.cornell.edu}
\altaffiltext{4}{CCPP, New York University, 4 Washington Place, New York, NY 10003; shan.huang@nyu.edu}
\altaffiltext{5}{Kapteyn Astronomical Institute, University of Groningen, Landleven 12, Groningen NL-9747AD, The Netherlands; papastergis@astro.rug.nl}

\begin{abstract}
\noindent

We investigate a sample of {3 dwarf elliptical galaxies} in the Virgo Cluster which have significant reservoirs of \HI. We present deep optical imaging (from CFHT and KPNO), \HI\ spectra (Arecibo) and resolved \HI\ imaging (VLA) of this sample. These observations confirm their \HI\ content and optical morphologies, and indicate that the gas is {unlikely to be} recently accreted. The sample has more in common with dwarf transitionals, although dwarf transitionals are generally lower in stellar mass and gas fraction. VCC 190 has an \HI\ tidal tail from a recent encounter with the massive spiral galaxy NGC 4224. In VCC 611, blue star-forming features are observed which were unseen by shallower SDSS imaging.


\end{abstract}

\section{Introduction}
\noindent
The relationship between gas-rich late-type dwarfs---irregulars and blue compact dwarfs (BCDs)---and early-type dwarfs---dwarf ellipticals and spheroidals---is unclear. Are late-type dwarfs the progenitors of early-types, or are the two populations largely distinct, with late-types evolving in the field and dwarf ellipticals in cluster environments? On one hand, many early-type dwarfs show late-type features: faint disk-like structures (\citealt{Lisker2006disk}; \citealt{Penny2014}), rotational support (\citealt{Beasley2009}; \citealt{Guerou2015}; \citealt{Toloba2016}), and lingering central star formation (\citealt{Lisker2006bc}; \citealt{DeLooze2013}). In turn, the underlying old stellar population of late-type dwarfs resemble dwarf ellipticals and spheroidals (\citealt{Meyer2014}). On the other hand, simulations (\citealt{Weinmann2011}; \citealt{Lisker2013}; \citealt{Mistani2016}) and stellar population modeling (\citealt{Rakos2004}; \citealt{Koleva2009}; \citealt{Paudel2010}; \citealt{Roediger2011}; \citealt{Weisz2014}; \citealt{Guerou2015}) suggest that early-type dwarfs essentially halted star formation at $z\sim1$, and so late-type dwarfs at $z=0$ are unlike their progenitors.

Essentially all dwarfs in the field are star-forming or in a starburst phase (\citealt{Lee2009}; \citealt{Geha2012}) while quiescent dwarf ellipticals predominate in the cluster (\citealt{Dressler1980}; \citealt{Binggeli1987}). If these populations are related, then it is assumed that the cluster environment is responsible for this transformation. As gas-rich late-type galaxies fall onto the cluster, their gas is removed via processes such as ram-pressure stripping (\citealt{Boselli2008}; \citealt{Boselli2016}) or galaxy harassment (\citealt{Mayer2001}; \citealt{AguerriGonzalez2009}). This process is relatively rapid, less than 100 Myr (\citealt{Boselli2008}). By the time a dwarf galaxy has crossed through the cluster core, they are almost entirely gas free ($>99$\% gas removed). Without gas, star formation ceases, the galaxy's colors redden, and it loses any irregular or spiral features, becoming a smooth dwarf elliptical.

Given that gas removal precedes morphological changes, there should be few, if any dwarf elliptical galaxies which still have a detectable reservoir of \HI. And indeed there are only a few, about 2\%, which do (e.g. \citealt{HuchtmeierRichter1986}; \citealt{diSeregoAlighieri2007}; \citealt{Taylor2012}). In the work of \citet{Hallenbeck2012}, hereafter H12, we identified a sample of 7 gas-bearing dwarf elliptical galaxies in the Virgo Cluster. The star formation in this sample is clearly suppressed: the galaxies are as red as typical dwarf ellipticals in the $g-i$, NUV$-r$, and FUV$-r$ bands. Additionally, their gas fractions ($GF\equiv M_\text{HI}/M_*$) are typical for unstripped dwarfs both in Virgo and the field \citep{Huang2012}. Based on this, we argued that the gas has been recently accreted.

How plausible is this assertion, given their evidence? Most dwarf ellipticals have cluster orbits which are highly radial \citep{Conselice2001}. This means that they will spend most of their time near the cluster edge---which is where the sample preferentially lies---where they can encounter clouds of neutral gas falling onto the cluster. {At their present positions outside the x-ray emitting region of Virgo, accretion is possible, as evaporation of clouds of neutral \HI\ due to the hot intracluster medium is relatively slow compared with the time it would take to accrete a cloud of \HI\ (H12).}


The purpose of this work is to test the re-accretion hypothesis of H12, using new deep optical and \HI\ observations. In \S\ref{sec:sample}, we review and update the sample of H12, removing galaxies whose \HI\ was not confirmed. In \S\ref{sec:results}, we present the results of new optical and \HI\ observations of the sample. We weigh the evidence for morphological transformation of the galaxies in \S\ref{sec:morphology}, for recent accretion in \S\ref{sec:accretion}, and for quenched star formation without gas removal in \S\ref{sec:starburst}. In \S\ref{sec:gasstripping}, we argue that VCC 190 has had gas removed via tidal stripping.

\section{Sample Selection and Updates}
\label{sec:sample}
\noindent
Our sample selection is outlined in Table 1, as well as below. In H12, we defined a sample of dwarf elliptical galaxies in the Virgo cluster. These galaxies were identified using the morphology and subcluster assignments  from the Virgo Cluster Catalog (VCC; \citealt{Binggeli1985}), updates from \citep{Binggeli1993}, and our own internal assignments. 

\begin{deluxetable}{lr}
\tablecolumns{2}
\tablewidth{0pt}
\tabletypesize{\scriptsize}
\tablecaption{Sample Counts}
\tablehead{
  \colhead{Selection Criteria} & \colhead{Galaxies}}
\startdata
Binggeli et al 1985 (VCC) Galaxies         & 2096\\
Dwarfs Ellipticals in Virgo                &  365\\
Red dEs with \HI\ Detections (H12)         &    7\\
\HI\ Confirmed by Follow-Up                &    5\\
Removal of VCC 956 and VCC 1993 (see text) &    3\\
\enddata
\tablecomments{\label{tab:sample-select}Number of galaxies in sample after each cut was applied. This work is concerned only with the 3 which remain after all cuts have been applied.}
\end{deluxetable}

Of these 365 dwarf ellipticals, 7 both had enough \HI\ such that they were detected in the {Arecibo Legacy Fast ALFA survey (\nolinebreak{ALFALFA}; \citealt{Giovanelli2005}; \citealt{Haynes2011})} and clearly lay along the red sequence (SDSS $g-r\geq0.45$).

\subsection{ALFALFA Follow-Up}
\noindent
The \HI\ detections of this sample were often near the detection limit of \nolinebreak{ALFALFA's}, with signal-to-noise ratios of $3<SN<10$. So, we re-observed the each object as part of a larger ALFALFA campaign to confirm the \HI\ in unusual and low signal-to-noise galaxies.

We performed a 3 minute ON/OFF observation of each galaxy using the single pixel L-Band Wide (LBW) receiver. These pointed observations have an rms of 1.1 mJy in 10 \kms\ wide channels, and are thus twice as sensitive as ALFALFA, which has a typical rms of 2.2 mJy when smoothed to the same width.

Out of the seven, the \HI\ emission in five of the galaxies was confirmed. The two galaxies with the lowest signal-to-noise (VCC 421 and VCC 1649) were not confirmed, and so we remove them from our sample.

\subsection{The Trouble With VCC 956 and VCC 1993}

\begin{figure}[!ht]
\epsscale{1.15}
\plotone{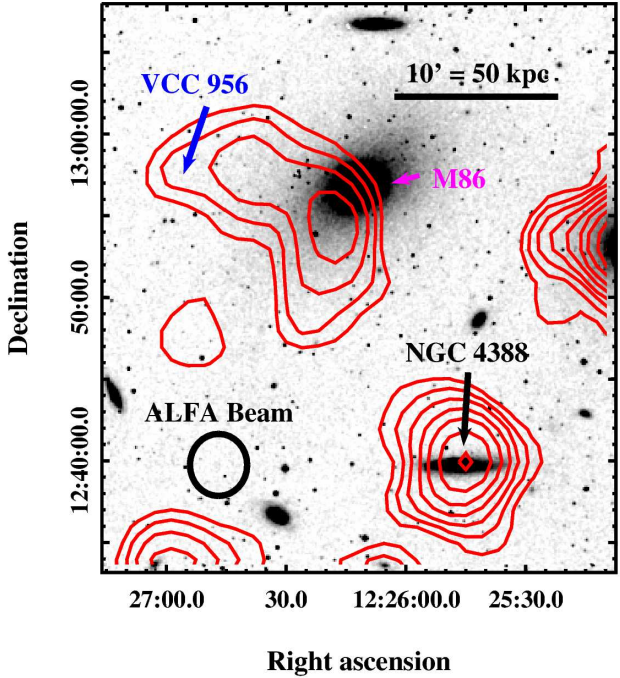}
\caption{VCC 956 and its local environment. VCC 956 lies along the tidal tail off of NGC 4388, and our \HI\ redshift coincides with the tail redshift of $2200$ \kms. It is thus likely that the \HI\ detected at the sky location of VCC 956 is not associated with the galaxy, and we remove it from our sample. Background is a DSS/II composite optical image, while contours are integrated \HI\ flux from ALFALFA covering $1820$ \kms $<v<2630$ \kms. Contours begin at $4\sigma$ (0.6 Jy km s$^{-1}$ beam$^{-1} = 0.1\ M_\odot$ pc$^{-2}$), and increase by a factor of $\sqrt{2}$ at each additional contour. The $3.8^{\prime}\times3.5^{\prime}$ ALFA beam is shown for reference.
\label{fig:vcc956-environment}}
\end{figure}

\citet{OosterloovanGorkom2005} mapped an \HI\ tail caused by ram-pressure stripping off of NGC 4388, which ranges in velocity from $2000<v<2600$ \kms. The sky position of VCC 956 coincides with the end of the tail, as shown in Figure \ref{fig:vcc956-environment}. Our nominal \HI\ detection has a velocity of $v=2200$ \kms, which perfectly coincides with the velocity of NGC 4388's tail at that position (see \citealt{OosterloovanGorkom2005}, Figures 1 and 2). There are no independent redshift measurements of VCC 956. It is thus likely that our nominal \HI\ measurement of VCC 956 is better identified with the tail of NGC 4388, and we remove it from our sample.


The surface brightness profile of VCC 1993 indicates that it is not a dwarf elliptical, but a low mass elliptical. It also has the highest $M_*$ in the sample by a factor of 10. So, we remove it from present consideration and focus on the lowest mass, truly dwarf, galaxies.

The remaining three galaxies, VCC 190, VCC 611, and VCC 1533 are the focus of this work. 
\\
\section{Observations and Results}
\label{sec:results}
\subsection{Summary of Prior SDSS, GALEX, and ALFALFA Observations}
\label{sec:summary}
\begin{figure*}[ht]
\begin{center}
\epsscale{0.57}
\plotone{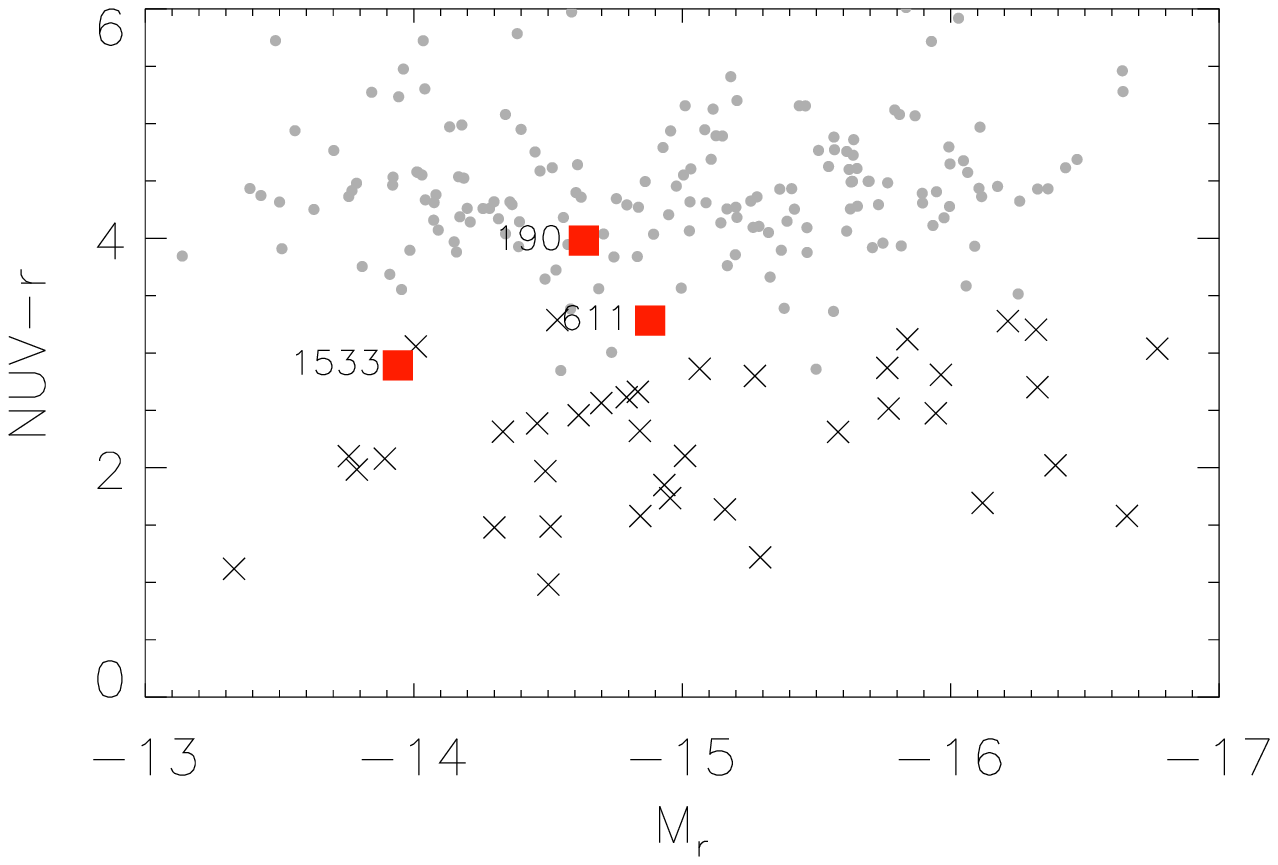}
\plotone{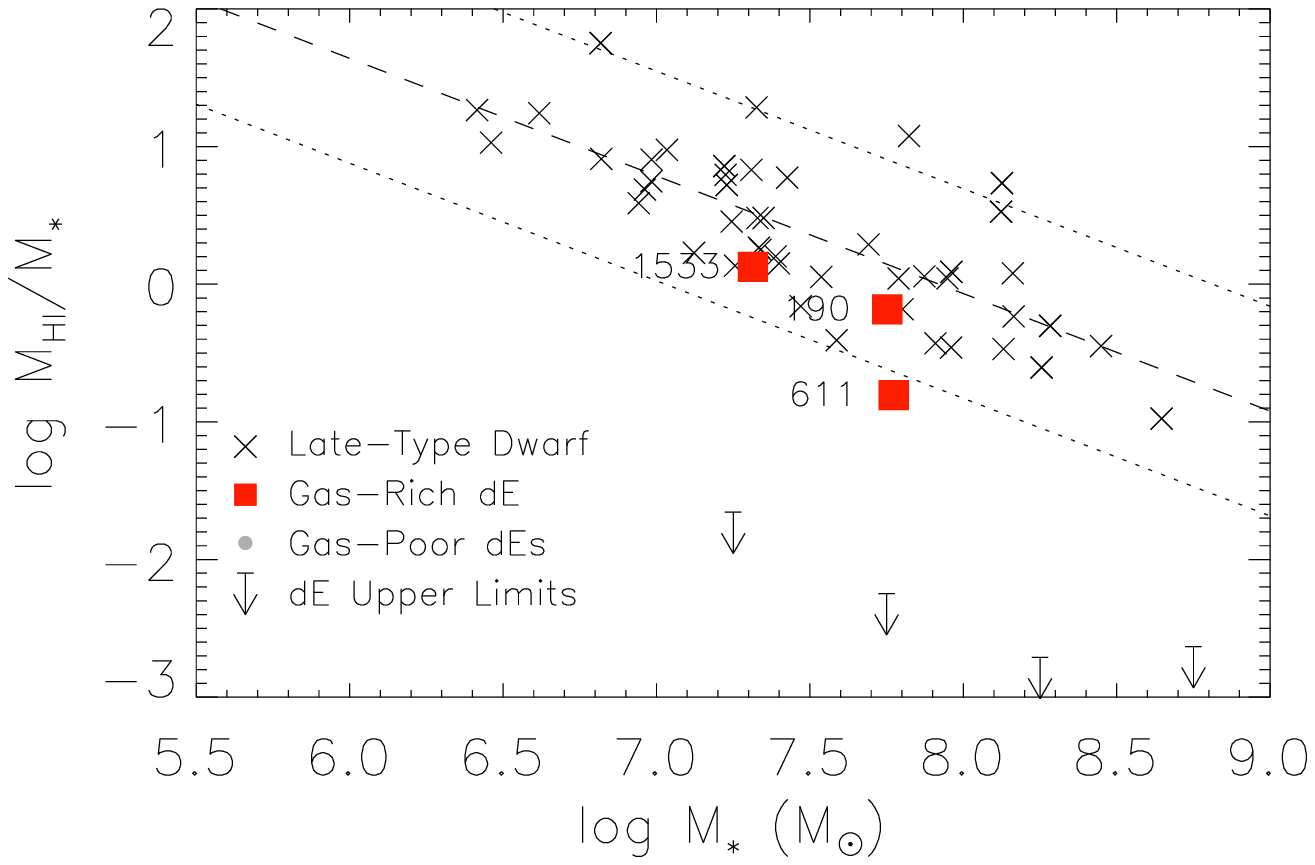}
\end{center}
\caption{The 3 dwarf galaxies in our sample are red like typical dwarf ellipticals yet similar in gas richness to star-forming dwarf irregulars in the Virgo Cluster. (left) Ultraviolet and optical color-magnitude diagram of the dwarf galaxies in the Virgo Cluster. Crosses are late-type dwarfs with gas, gray circles are dwarf ellipticals without gas. Our sample of gas-bearing dwarf ellipticals are labeled with red squares. (right) Gas fraction as a function of stellar mass for late type dwarfs and our sample. Lines indicate the best fit and $2\sigma$ scatter for the late type dwarfs, and arrows are the $3\sigma$ upper limit for non-detected dwarf ellipticals based on stacking. All measurements are from \citet{Huang2012} with updates in H12.
\label{fig:colors}}
\end{figure*}

\noindent
Figure~\ref{fig:colors} (left) shows an NUV-r color-magnitude diagram of the dwarf sample of H12. Late-type dwarfs with \HI\ are crosses, while early type dwarfs undetected in ALFALFA are gray dots. The three gas-rich dwarf ellipticals in our sample are plotted as red squares, and labelled with their VCC numbers. With a few exceptions, the late-type and early type dwarfs separate themselves neatly. From our sample, VCC 190 is unambiguously red in NUV$-r$ color. VCC 611 and VCC 1533 lie in the green valley; while they are redder than almost all of the late-type dwarfs, they are also bluer than most dwarf ellipticals. The galaxies show a similar segregation in FUV$-r$ and $g-r$ colors (see H12; Figure~6).

Figure~\ref{fig:colors} (right) shows the \HI\ gas fraction as a function of stellar mass of late type dwarfs (crosses) along with a best-fit trend. The \HI-bearing late-type dwarfs lie along the same line as \HI-bearing dwarfs outside of the Virgo cluster (H12; \citealt{Huang2012}). This suggests that such galaxies have yet to undergo any significant stripping processes. Red boxes with labels are our sample. All three galaxies lie at or below the line defined by the HI-bearing late types, but none significantly so. 

Arrows are 3$\sigma$ upper limits on the gas fractions of dwarf ellipticals which were not detected in ALFALFA derived from spectral stacking. The Arecibo Galaxy Environment Survey (AGES; \citealt{Auld2006}) found similar upper limits over a smaller region of the cluster (\citealt{Taylor2012}). 

The three dwarf ellipticals in our sample thus fall into an unusual class: from their optical and ultraviolet magnitudes alone, they appear to fit in with the dwarf elliptical population by being either ``red and dead'' (VCC 190), or nearly so (VCC 611 and 1533). However, from their \HI\ gas fractions, they appear to be unstripped late-type dwarfs, and several orders of magnitude more gas-rich than other dwarf ellipticals.

\subsection{Optical Observations}
\label{sec:optical}
\noindent
We obtained optical r-band imaging at the WIYN 0.9m telescope with the HDI camera at the Kitt Peak National Observatory (KPNO). VCC 190 and VCC 611 were observed for 30 and 36 minutes, respectively, with seeing $1^{\prime\prime}.6$. In addition, all three galaxies lay within the footprint of the Next Generation Virgo Survey (NGVS; \citealt{Ferrarese2012}), performed with the CFHT. All three galaxies were observed for between 30 minutes and 1 hour in the $u$, $g$, and $i$ bands.

\begin{figure*}[!htp]
\epsscale{1.1}
\plotone{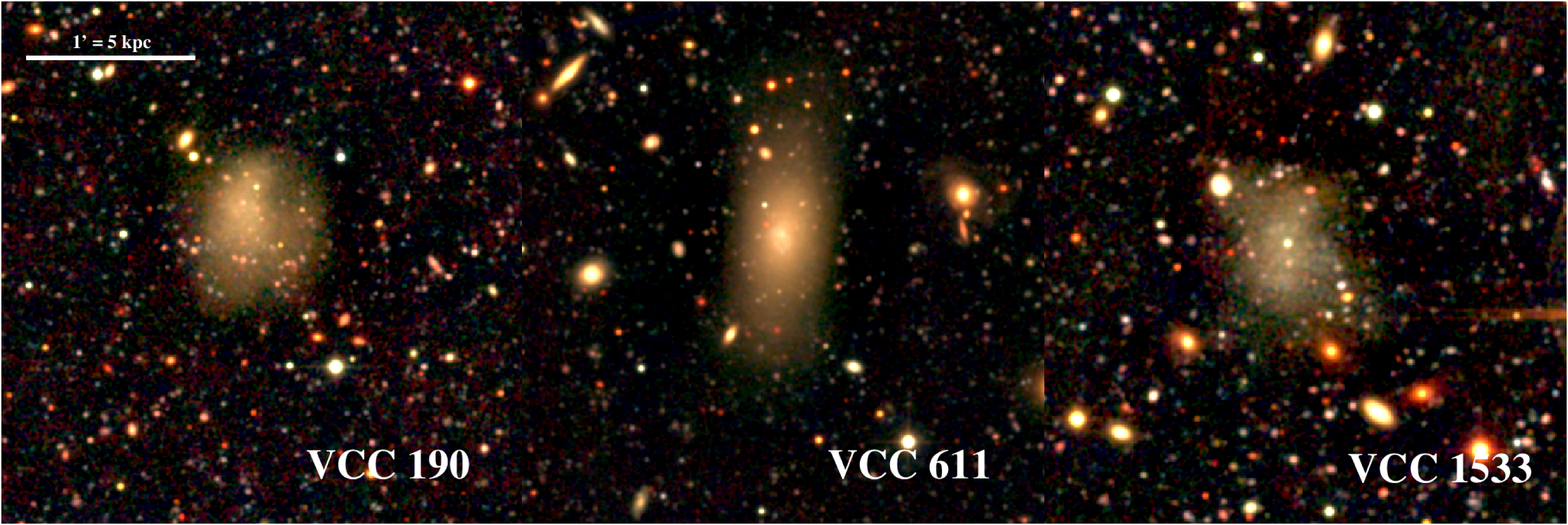}\\
\plotone{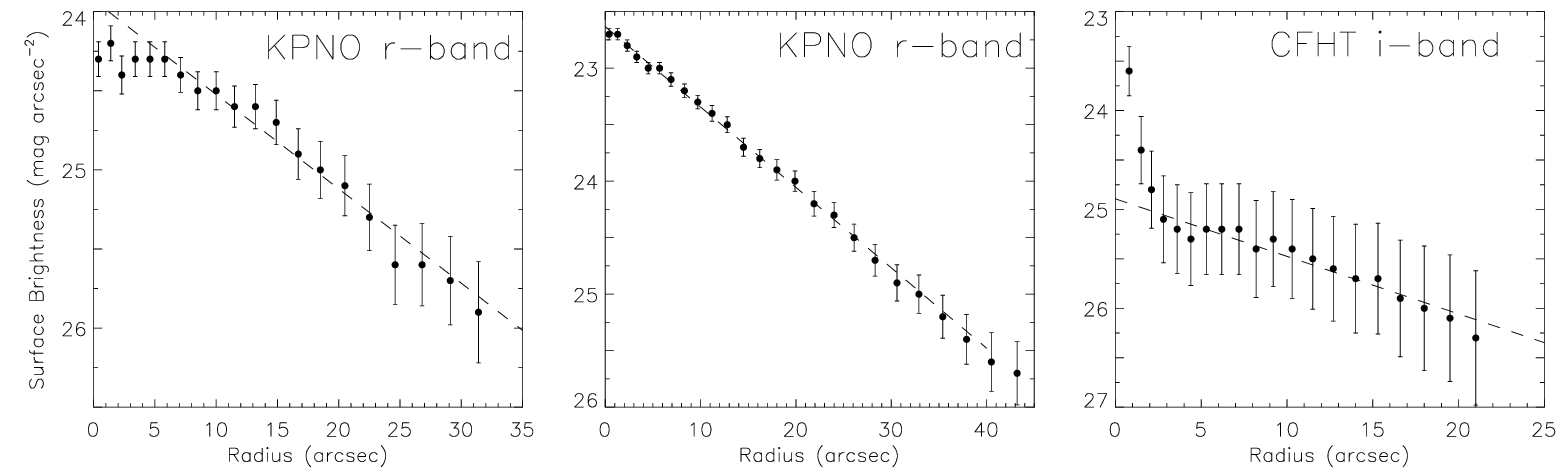}\\
\plotone{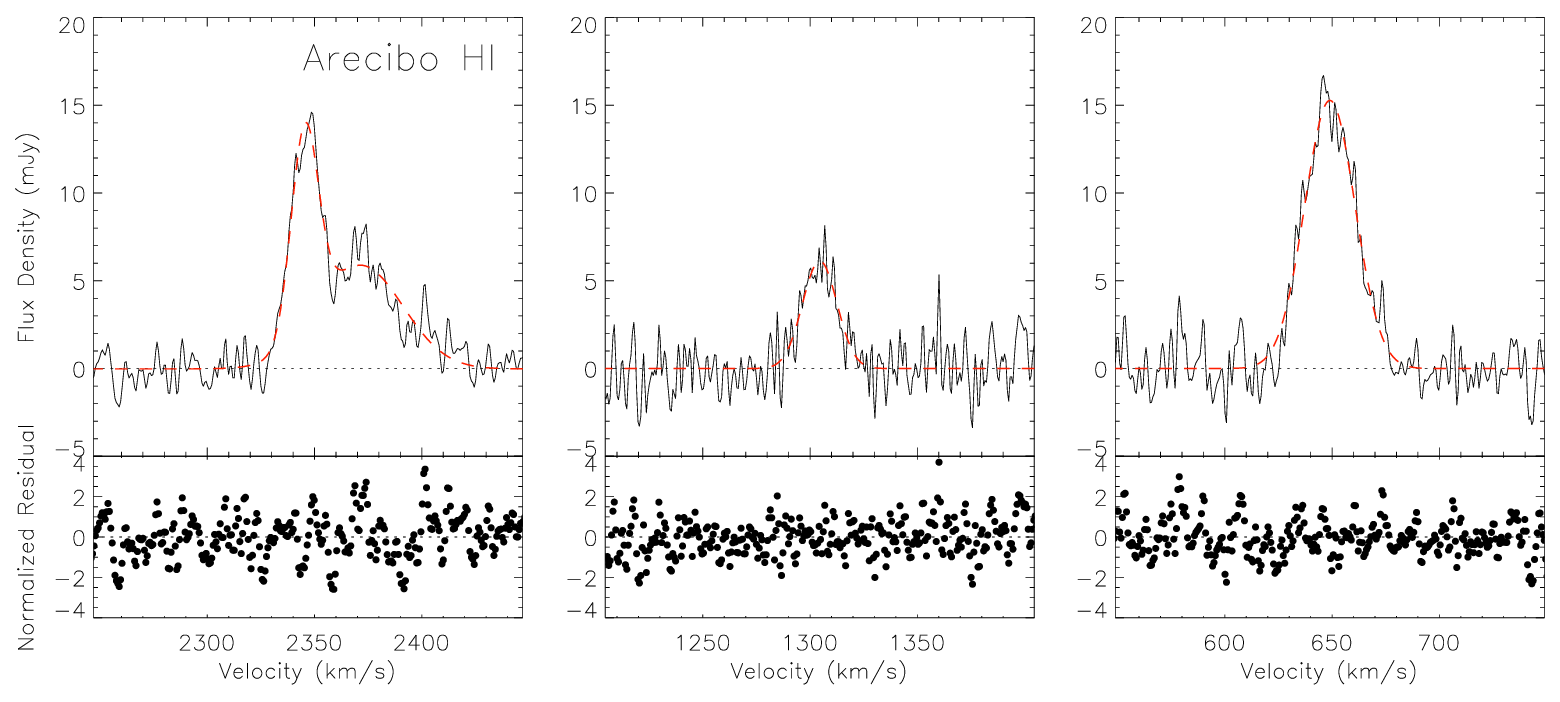}
\caption{Optical images and \HI\ spectra of our sample of `red and dead' yet gas-bearing dwarf ellipticals. (top row) Optical CFHT images of the three red dwarf elliptical galaxies in the sample, using $u$-, $g$-, and $i$-band filters. (second row) Surface brightness profiles of the three galaxies; dashed lines are exponential profiles fit to $r > 5^{\prime\prime}$. (third row) \HI\ Spectra; red dashed lines are best-fit gaussians profiles, except for VCC 190, where two gaussians are fit. (bottom) Normalized residuals from the gaussian fits. The lack of trend demonstrates that all three galaxies are well described by the fits.\label{fig:sample}}
\end{figure*}

Optical NGVS images of all three galaxies in our sample are shown in Figure~\ref{fig:sample} (top panels). The three galaxies all appear roughly elliptical but with some peculiar morphology. The highest optical surface density in VCC 190 (top left) is offset from the center, with possible tails to the west and south. The shape of VCC 1533 (right) is boxy, with a nucleated center. Figure~\ref{fig:sample} (second row) shows r-band (VCC 190 and 611) and i-band (VCC 1533) surface brightness profiles. All three galaxies are well-fit with an exponential profile for $r > 5^{\prime\prime}$ (dashed line), consistent with the standard dwarf elliptical profile. At smaller radii, the surface brightness profiles of VCC 1533 is contaminated by what is likely a foreground star.

\begin{figure}[ht]
\epsscale{1.1}
\plotone{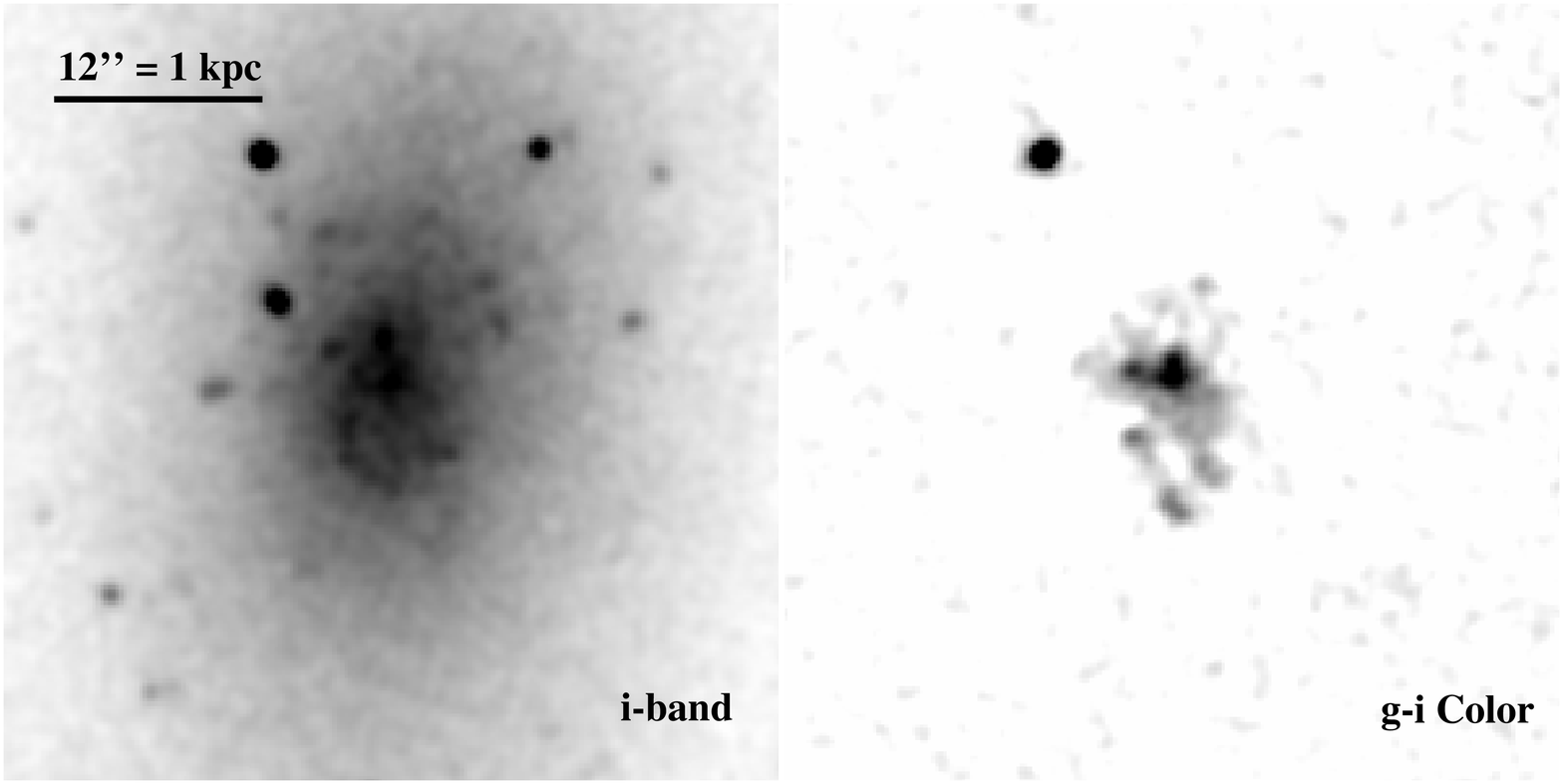}
\caption{Deep CFHT images reveal that VCC 611 has strong star formation concentrated at the galaxy's center. (left) CFHT $i$-band image of the central 4 kpc of VCC 611. (right) $g-i$ color of the same field. The $i$-band image shows a mostly smooth old stellar population, while the $g-i$ color shows that star formation is occurring in the galaxy's center, in several star-forming knots.\label{fig:bluecenter}}
\end{figure}

VCC 611 (top center) appears like a smooth elliptical, likely due to its relatively higher stellar mass and surface brightness compared with the other two galaxies in our sample. However, it clearly has a very blue center indicative of star formation, which can be seen in Figure~\ref{fig:bluecenter}. The left panel presents an $i$-band image, which shows a smooth old stellar population. The right panel is a $g-i$ color map of the galaxy, which shows several star-forming knots in the galaxy's center. VCC 611 was observed as part of the SDSS spectroscopic survey and has \Halpha\ in emission with an equivalent width (EW) of 25 \AA \ (NASA Sloan Atlas\footnote{www.nsatlas.org}). \citet{Lisker2006disk} and \citet{Taylor2012} previously performed unsharp masking on all three of these galaxies using shallower SDSS images, but did not detect the underlying structure or the blue center in VCC 611.\\\\

\subsection{HI Observations}
\label{sec:HI}
\noindent
The HI content of the three dwarf elliptical galaxies in our sample were observed at high spectral resolution (native 0.65 \kms\ channel width) using the Arecibo Observatory. Observations were performed with the LBW receiver in a series of 5 minute ON/OFF pairs. Results are summarized in Table 2. The ALFALFA data cubes indicate that there are no sources within $10^\prime$ which could be contaminating the sources by being detected in Arecibo's sidelobes above the 5\% level.

The \HI\ spectra of the three galaxies are shown in Figure~\ref{fig:sample} (third row). VCC 611 and VCC 1533 (center and right) are in good agreement with a gaussian profile: the normalized residuals (bottom center and right) show low scatter and non-significant trends. In addition, we fit 3rd order gauss-hermite polynomials, to both spectra to look for asymmetric components. For VCC 611, we find no significant asymmetries, while VCC 1533's \HI\ profile is asymmetric at a $3\sigma$ level. The velocity widths of both galaxies are very narrow: $W_{50} = 19$ and $28$ \kms, respectively). Fit parameters are presented in Table 3.

\begin{deluxetable}{lccccc}
\tablecolumns{5}
\tablewidth{0pt}
\tabletypesize{\scriptsize}
\tablecaption{Sample HI Properties}
\tablehead{
  \colhead{Galaxy} & \colhead{$S_\text{HI}$} & \colhead{$\log M_\text{HI}$} & \colhead{$V_{sys}$}       & \colhead{$W_{50}$}\\ 
                   & Jy \kms                 & $\log M_\odot$               & \kms                      & \kms\\
                   & (1)                     & (2)                          & (3)                       & (4)}
\startdata
VCC 190 A          & $0.18 \pm 0.01$         & 7.1                          & $2345.4 \pm 0.2$          & $14.9 \pm 0.6$\\
VCC 190 B          & $0.29 \pm 0.02$         & 7.3                          & $2371$ \ \ $\pm$ $2$ \ \  & $46 \pm2$\\
VCC 611            & $0.12 \pm 0.01$         & 6.9                          & $1304.5 \pm 0.6$          & $19 \pm 1$\\
VCC 1533           & $0.45 \pm 0.01$         & 7.5                          & \ $648$ \ \ $\pm$ $1$ \ \ & \ $27.7 \pm 0.6$
\enddata
\tablecomments{\label{tab:spectra}Results of single pixel \HI\ spectral observations. VCC 190 is split into a relatively blueshifted (A) and redshifted (B) peak. Column 1: integrated flux density; Column 2: \HI\ mass, assuming a distance of 16.7 Mpc; Column 3: heliocentric optical velocity; Column 4: Full-width half maximum of gaussian fit. For VCC 190, the fit is two simultaneous gaussian fits, one to the narrower, low velocity piece, and to the wider, higher velocity tail.}
\end{deluxetable}

The \HI\ profile of VCC 190 (Figure~\ref{fig:sample}; third row left) is neither gaussian nor does it have a typical symmetric two-horn profile. Attempting to fit a single, asymmetric profile to the galaxy produces a rather poor fit overall ($\chi^2_\nu = 3.0$). Such asymmetries in the global \HI\ profile are typical in galaxies for which ongoing ram-pressure stripping is observed (\citealt{KoopmannKenney2004}; \citealt{OosterloovanGorkom2005}; \citealt{Haynes2007}; \citealt{Chung2007}), though stripping is not the only possible explanation.

Instead, guided by our VLA observations (see \S\ref{sec:vla}), we fit two gaussians, one to the lower redshift peak, and one to the higher redshift peak. The combined model fits the data well ($\chi^2_\nu=1.5$), with no trend in the residuals (bottom left). The lower redshift peak has a very narrow $W_{50} = 15$ \kms, while the higher redshift peak has $W_{50} = 46$ \kms. 

\subsection{VLA Observations of VCC 190}
\label{sec:vla}
\noindent
We observed VCC 190 using the VLA in the D and C configurations for 2 hours and 4 hours respectively. The cubes were produced in CASA using multiscale clean, with a Briggs robustness weighting of $1.0$. Continuum subtraction was performed in the image plane with the \textsc{imcontsub} task using the line-free channels. This yielded an rms of $1.0$ mJy \perbeam\ in a $37^{\prime\prime}\times 31^{\prime\prime}$ beam, with  $7$ \kms\ wide channels. The moment 0 maps were produced by first smoothing the data cubes to half their original spatial resolution and calculating a mask at 3 $\sigma$ (3.0 mJy \perbeam). This mask was then applied to the original data cube.

\begin{figure*}[!ht]
\begin{center}
\epsscale{1.0}
\plotone{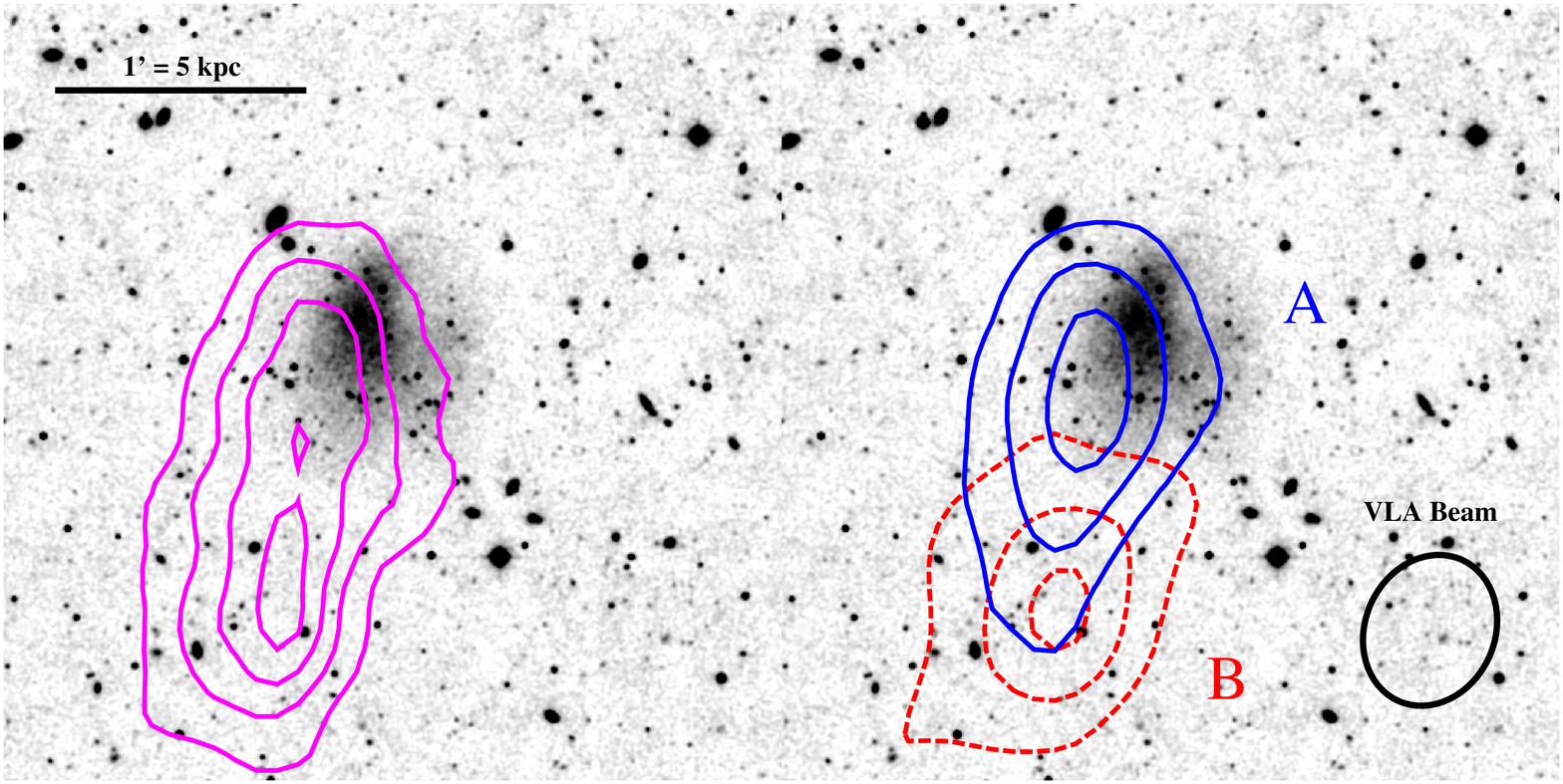}
\end{center}
\caption{VLA \HI\ intensity maps of VCC 190 overlaid on a KPNO $r$-band image. There are two separate reservoirs of gas in VCC 190, an undisturbed component corresponding roughly to the optical center of the galaxy, and a wide ($W_{50} = 50$ \kms) component corresponding to the tail. Contours begin at $3\sigma$ (0.03 Jy \kms\ beam$^{-1} = 0.2\  M_\odot$ pc$^{-2}$) and increase by $3\sigma$ at each additional contour. (left) Intensity map including gas at all velocities (right) Because the galaxy has two distince peaks, we produce two overlaid maps, one corresponding to the blueshifted peak ($v < 2343$ \kms; blue solid contours) and the other for the redshifted velocity peak ($v > 2343$ \kms; red dashed contours). The lower velocity peak corresponds to gas coincident with the optical disk of the galaxy {(VCC 190 A)} while the higher velocity peak corresponds to an \HI\ tail {(VCC 190 B)}.
\label{fig:vcc190-vla}}
\end{figure*}

Figure~\ref{fig:vcc190-vla} (left) shows a moment 0 image overlaid on a KPNO $r$-band optical image. Contours begin at $3\sigma$ and increase by $3\sigma$ at each additional contour. The gas is not localized directly on the optical, but shows a tail to the southeast. The total flux recovered by the VLA is $0.44\pm0.01$ Jy km s$^{-1}$, which agrees with the Arecibo flux of $0.47 \pm 0.02$ Jy km s$^{-1}$. In Figure~\ref{fig:vcc190-vla} (right), we overlay two sets of contours, corresponding to the velocities of the two peaks observed at Arecibo. The blue contours, corresponding to the narrow, lower velocity peak ($2320-2343$ \kms) VCC 190 A roughly coincide with the location of the optical galaxy, with an offset of $12^{\prime\prime}=1$ kpc. The red dashed contours correspond to the wider, high velocity peak ($2343-2374$ \kms) VCC 190 B correspond solely to the tail.

\section{Galaxy Morphology}
\label{sec:morphology}
\noindent
In H12, we argued that this sample is morphologically well-described as dwarf ellipticals, based on optical morphology and color from relatively shallow imaging. All three galaxies in our refined sample were classified as dwarf ellipticals by \citet{Binggeli1985}, are redder than almost all gas-rich late-type dwarfs in the cluster, and no faint substructure was recovered in the galaxies by either \citet{Lisker2006disk} or \citet{Taylor2012} using stacked SDSS imaging. Do the new, significantly deeper, optical and \HI\ observations support this assessment?

In the CFHT images, we observe that only VCC 611 appears by eye to be a `typical' dwarf elliptical: it is smooth and clearly elliptically symmetric. VCC 190 and 1533 are a bit peculiar (see Figure~\ref{fig:sample}, first row): VCC 190 is asymmetric and VCC 1533 is boxy in shape. Such peculiar features are, however, not unusual. Massive elliptical galaxies in the field which are observed to have detectable quantities of \HI\ often have peculiar optical morphology, appearing blue, distorted, or with faint disk-like features (e.g. \citealt{diSeregoAlighieri2007}; \citealt{Hau2008}; \citealt{Grossi2009}; \citealt{Serra2012}). 

It is possible that VCC 611 is a massive elliptical galaxy, but appears to be a dwarf because it is behind the cluster. This can be tested using surface brightness profiles: all dwarf galaxies (early- and late-type) are expected to have exponential profiles, as compared with the de Vaucouleurs profile of more massive ellipticals.  Using $r$- and $i$-band imaging from KPNO and the CFHT, we observe that exponential profiles do fit all three galaxies very well at radii larger than $5^{\prime\prime}$ (see Figure~\ref{fig:sample}, second row). This measure does not, unfortunately, allow us to differentiate between dwarf ellipticals and irregulars.

The bulk motion of stars and gas in dwarf ellipticals is generally not rotational, but pressure-supported. Typical values of $V_{rot} \lesssim \sigma$, where $\sigma$ is the dispersion velocity. For the stellar component, typical values are $0$ \kms $<V_{rot}<30$ \kms, while $20$ \kms $< \sigma_*<40$ \kms\ (e.g. \citealt{vanZee2004}; \citealt{Rys2013}; \citealt{Toloba2014}). This relationship only seems to break down for the most extreme of the so-called fast-rotating and rotationally supported dwarf ellipticals (\citealt{Beasley2009}).

In agreement with stellar observations of other dwarf ellipticals, all three galaxies in our sample have \HI\ profiles which show little sign of rotation. All three can be very well fit with a gaussian profile, with widths of $14$ \kms $< W_{50} < 28$ \kms\ (see Figure~\ref{fig:sample}, bottom two rows). We can estimate a sky-projected rotation velocity via:

\begin{equation}
V_{rot}=\frac{\sqrt{W_{50}^2-\sigma_\text{HI}^2}}{2}
\end{equation}

\noindent where $\sigma_\text{HI} = 11$ \kms\ is our assumed velocity dispersion, and the factor of two accounts for the integrated spectrum containing both the approaching and receding half of the rotation curve. We then obtain approximate \HI\ rotational velocities of $4-13$ \kms, or $V_{rot} \lesssim \sigma_\text{HI}$.

By comparison, the estimated rotational velocities for other late-type dwarfs is much higher. For late-types dwarfs in the Virgo Cluster observed by ALFALFA, the average rotational velocity is $31$ \kms, with an interquartile range (IQR) of $18$ \kms $< V_{rot} < 40$ \kms, several times $\sigma_\text{HI}$. The FIGGS sample of field irregulars \citep{Begum2008} are typically narrower, with an average $V_{rot}$ of $39$ \kms (IQR of $11$ \kms $< V_{rot} < 24$ \kms. While the rotational velocities of the FIGGS sample just overlaps with our sample, the FIGGS galaxies are also typically of lower mass. Finally, we note that the spectra of most late-type dwarfs in ALFALFA and FIGGS also have a two-horned profile or otherwise show clear features of rotational broadening.


\section{Discussion}
\noindent
Thus, our previous assessment that these galaxies are morphologically dwarf ellipticals appears to hold, regardless of whether we consider surface brightness profiles, visual appearance, or velocity widths.
There are, however, a few peculiarities which must be explained, primarily their large \HI\ reservoirs in comparison with other dwarf ellipticals. We now explore two hypotheses about the \HI\ content of the galaxies. First, we reconsider the hypothesis of H12 that the gas has been recently accreted. Second, we consider whether these galaxies are a part of the irregular or BCD dwarf population, but are between starburst phases, and so only appear optically red and dead.

\subsection{Gas Accretion}
\label{sec:accretion}
\noindent
{Approximately half of giant elliptical galaxies in the field have been observed to bear \HI\ (\citealt{Morganti2006}; \citealt{Hau2008}; \citealt{Grossi2009}; \citealt{Oosterloo2010}; \citealt{Serra2012}). In most cases, these authors attributed the presence of gas to accretion---whether of gas-rich satellites, the cooling of ionized gas, or cold-mode accretion from the intergalactic medium (e.g. \citealt{Struve2010}; \citealt{Davis2011}; \citealt{Serra2012}; \citealt{DeRijcke2013}). However, in clusters, ellipticals do not appear to be accreting \citep{Oosterloo2010}. \citet{Morganti2006}, \citet{Oosterloo2010}, and \citet{Serra2012} observed the \HI\ in a combined 54 gas-bearing giant ellipticals in the field and the Virgo cluster. These authors found a wide range of \HI\ morphologies, from scattered clouds and tails to regularly rotating \HI\ disks. They infer recent or ongoing accretion as the origin of the gas when the \HI\ is in the form of clouds, tails, and warped and disturbed disks. \citet{Oosterloo2010} also observed that star formation is generally observed in galaxies where accretion is recent or ongoing. When accretion is not recent, star formation is lacking, even when a galaxy is gas-rich.}

{However, the most appropriate population to compare our sample to is other low-mass early-type galaxies, such as NGC 404 and FCC046 (\citealt{delRio2004}; \citealt{DeRijcke2013}). \HI\ has been observed in both galaxies, and attributed by the authors to gas accretion in the last Gyr. Like the more massive ellipticals, the gas in NGC 404 and FCC046 is rotating in a disk. The rotational structure of the gas is clear even in an integrated spectrum: both galaxies have a classic two-horned profile, with a velocity widths of 65 \kms\ \citep{delRio2004} and 67 \kms\ \citep{DeRijcke2013}.\footnote{We have re-fit the spectrum of FCC046, as the $W_{50} = 34$ \kms\ \citet{DeRijcke2013} report is from a gaussian, not two-horned, fit to the spectrum.} We note that the true rotational velocity for NGC 404 must be quite large, as its \HI\ disk is observed nearly face-on. For FCC046, the misalignment between the angular momentum of the accreted gas and the stellar component of the galaxy is very clear: the gas is in a polar ring. Like more massive ellipticals, accretion coincides with recent star formation for these dwarfs (\citealt{Thilker2010}; \citealt{DeRijcke2013}).}

{Whether comparing with massive or dwarf ellipticals, it is clear that accreted gas is associated with two features: the \HI\ morphology is in the form of gas clouds, streams, or warped or disturbed disks, and the presence of star formation. For our sample---where our primary observations are single dish spectra---the direct observable of \HI\ in streams or disks would be a large velocity width ($\gtrsim 50$ \kms) or a two-horned profile as a signature of a rotating disk. We note however that the spectral features are not a perfect analogue to resolved \HI\ morphology and are insufficient to conclusively point to recent accretion as the source of the gas. Most notably, we would be unable to discern the existence of a warp in an otherwise smoothly rotating disk. Instead, we can argue that the lack of such features weakens the case for accretion. Neither wide velocity widths nor two-horned profiles are observed for any of the three galaxies. We do, however, observe evidence of recent star formation in VCC 611 and VCC 1533. For VCC 190, where the \HI\ is partially resolved, we do observe a gas tail, but we argue that gas removal is a more plausible explanation (see \S\ref{sec:gasstripping}).}

{Finally, we consider the properties of free \HI\ clouds in the Virgo cluster. Such clouds would be the source of the accreted \HI\ in our sample. Here we run into a signficant problem. If the accretion is ongoing or recent, then a tidal interaction between the clouds and the galaxies would increase the velocity widths of the clouds' \HI\ profiles. However, the velocity profile of the \HI\ in each of the three galaxies is much narrower than that of any of the clouds in Virgo, which have $W_{50} \gtrsim 50$ \kms\ (\citealt{Kent2007}; \citealt{Kent2009}). }

\subsection{Between Bursts of Star Formation}
\label{sec:starburst}
\noindent
The high gas fractions in our sample (see Figure~\ref{fig:colors}, right) and their location at the edge of the cluster (\citealt{Conselice2001}; \citealt{Cen2014}; \citealt{Jaffe2016}) suggest that these galaxies are new arrivals to the cluster. However, essentially all dwarf galaxies in the field show signs of recent star formation, to varying degrees (\citealt{Lee2009}; \citealt{Geha2012}). If these galaxies were quenched prior to arriving in the cluster, then approximately 50\% of their gas reservoirs could have been removed, consistent with \HI\ observations (\citealt{Hess2013}; \citealt{Odekon2016}) and simulations (\citealt{Bekki2009}; \citealt{Taranu2014}) of galaxy groups.

In groups, dwarfs with morphological features which are somewhere between irregulars and spheroidal dwarfs are observed (\citealt{Mateo1998}; \citealt{SandageHoffman1991}). We observe such features in our sample: VCC 190 and VCC 1533 show deviations from a purely elliptical shape, and VCC 611 has some star-forming features at its center. Called dwarf transitionals, such galaxies have little ongoing star formation, and often have detectable reservoirs of \HI. The star formation history of such galaxies is very similar to irregulars and spheroidals until roughly 1 Gyr ago (\citealt{Weisz2011a}; \citealt{Weisz2011b}). 
The general interpretation is that dwarf transitionals are either between major episodes of star formation, or have simply permanently stopped forming stars (\citealt{Grebel2003}; \citealt{Skillman2003}).

The typical dwarf transitional has a lower $M_*$ and much lower GF than our sample (\citealt{Weisz2011a}; \citealt{DeLooze2013}). Our sample could be the gas-rich and more massive tail of the dwarf transitional population, holding on to more of their gas while in the group environment by virtue of their higher total mass. Indeed, our most gas-poor galaxy, VCC 611, appears in the dwarf transitional sample of \citet{DeLooze2013}.







\section{The Tail of VCC 190}
\label{sec:gasstripping}
\noindent
Of the three galaxies in our sample, only VCC 190 has clear ongoing gas removal, as indicated by both its unusual HI spectrum and tail. The two processes most likely to produce such a tail are ram pressure stripping and a tidal interaction with a larger galaxy or galaxies. Which is it?

Evidence for ram pressure stripping is fairly weak. VCC 190's projected position is beyond the $3\sigma$ detection of the ROSAT satellite \citep{Bohringer1994}, where the effects of the ICM should be minimal. In addition, massive galaxies at this distance from the cluster center are not \HI\ deficient, and rarely show ram pressure stripping tails \citep{Chung2007}. However, VCC 190 is very low mass, and may be susceptible to ram pressure stripping even when it is too weak for the more massive galaxies.

The tail geometry is also not right for ram pressure stripping. In general, tails point opposite the motion of the galaxy, which generally means they point roughly away from the subcluster center (e.g. \citealt{Chung2007}; \citealt{Roediger2008}; \citealt{Kenney2014}). In this case, the tail should point away from either M87 or M49, to the northwest and west, respectively, but VCC 190's tail points southwest. In addition, tails caused by stripping generally point towards the mean redshift of the cluster, as the removed gas is decelerated relative to the ICM, but VCC 190's tail is blueshifted towards higher velocity.

Instead, the tidal interaction hypothesis is stronger. First, we note that the distribution of stars in VCC 190 is not a smooth ellipse, but asymmetric; the top left portion of the galaxy has a higher surface brightness than the bottom right. Ram pressure stripping would only disturb the gas in the galaxy, while a strong tidal encounter, or multiple weaker encounters can disturb the stars as well.

\begin{figure}[!ht]
\begin{center}
\epsscale{1.15}
\plotone{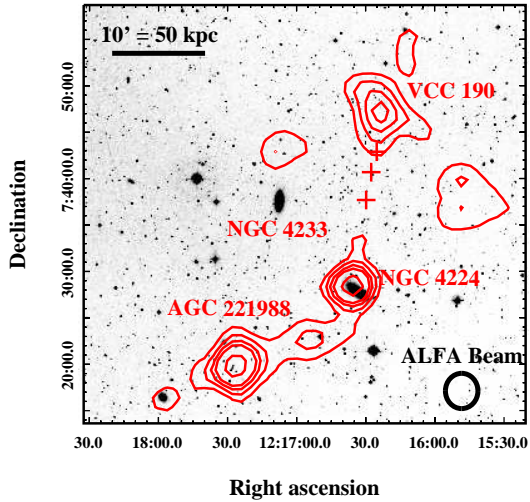}
\end{center}
\caption{The environment of VCC 190. The tidal tail of VCC 190 points towards NGC 4224, suggesting a tidal encounter between the two. NGC 4224 is itself interacting with AGC 221988. Background is a DSS-2 composite optical image, while the overlaid contours are an \HI\ column density map; contour levels are identical to Figure \ref{fig:vcc956-environment}. All galaxies with redshifts consistent with the Virgo Cluster are labeled. The red crosses are the locations of additional Arecibo pointings used to detect a bridge between NGC 4224 and VCC 190; no gas was detected at any of the three pointings.
\label{fig:vcc190-environment}}
\end{figure}

Second, it is clear which galaxies collided with VCC 190. With the exception of a few dwarf galaxies, three galaxies are within 100 kpc of VCC 190 and at a similar redshift, all to the southwest. Figure~\ref{fig:vcc190-environment} shows an optical image with ALFALFA contours overlaid of these galaxies. NGC 4224 and AGC 221988 are interacting, with a tidal bridge visible between them (see Figure~\ref{fig:vcc190-environment}). \citet{Miskolczi2011} found stellar tidal streams coming off of NGC 4224 pointing to the north, for which they could not find an obvious interaction partner; we suggest that AGC 221988 is the cause. An optical spectrum indicates that NGC 4233 is at the same redshift, and so may be interacting with the other galaxies. But, its lack of \HI\ means that such an interaction unseen in the image.

The tidal tail of VCC 190 points toward NGC 4224, suggesting an encounter between the two. In addition, NGC 4224 lies at a slightly higher redshift ($v = 2606$ \kms) than VCC 190 ($v = 2345$ \kms). As VCC 190 passed by NGC 4224, the removed gas in VCC 190's tail would be perturbed to a higher redshift, as is observed. We performed additional observations at Arecibo between VCC 190 and NGC 4224 (Figure~\ref{fig:vcc190-environment}; red crosses) to look for a tidal bridge between the two galaxies. No gas was detected, with a $3\sigma$ upper limit of $10^{6.3} M_\odot$ (0.10 Jy \kms), assuming a $30$ \kms\ wide tail, or approximately 5\% of the \HI\ mass of VCC 190.


Based on the disturbed and asymmetric stellar light distribution in the galaxy, the presence of a narrow and undisturbed ($W_{50} = 14$ \kms) component of HI centered on the optical galaxy, and finally NGC 4224 as a clear collision partner, this tail is likely to be tidal in nature.

\section{Summary}
\label{sec:conclusions}
\noindent
We revisited sample of dwarf elliptical galaxies in the Virgo Cluster identified by H12. Despite being `red and dead,' all had substantial reservoirs of \HI\ as detected in ALFALFA. We found that the two lowest signal-to-noise \HI\ detections reported in H12 and \citet{Haynes2011}, VCC 421 and VCC 1649, were spurious. We further eliminated two more galaxies from the sample of H12: VCC 956 appears at the same sky location as a long \HI\ tidal tail off of NGC 4388, and VCC 1993 is not a dwarf galaxy. Our primary sample of interest was thus narrowed to three galaxies: VCC 190, VCC 611, and VCC 1533.\\
\\
\textbf{New radio and optical observations.} High resolution (channel width of 0.65 \kms) Arecibo observations reveal that VCC 611 and 1533 have narrow ($W_{50}<30$ \kms) velocity widths and gaussian profiles. VCC 190 has two gaussian peaks, one narrow and one wide ($W_{50}=45$ \kms). VLA observations of VCC 190 show an \HI\ tail pointing toward NGC 4224 to the south, and the tail corresponds to the wide velocity feature observed at Arecibo. The ALFALFA data cubes show that NGC 4224 is interacting with one of its neighbors, AGC 221988.

Our KPNO observations and CFHT archival data from the NGVS show that all three galaxies have exponential profiles and lack widespread star-forming features. They do have some unusual features, however: the surface brightness of VCC 190 is asymmetric, VCC 611 has a star-forming feature at its center, and VCC 1533 is boxy in shape.\\
\\
\textbf{Elliptical morphology without significant removal of gas} is observed for all three galaxies, even though gas removal in clusters should be very rapid compared with morphological change. This is supported both by their optical (exponential profiles and a lack of widespread star-forming features) and \HI\ (narrow, undisturbed velocity profiles) features. Together, these features suggest that the galaxies are relatively new arrivals in the cluster which have not {undergone significant gas stripping. We observe ongoing gas removal in VCC 190 (see below), but the current tidal interaction is too recent to be responsible for its optical properties.}

These galaxies are similar to dwarf transitionals---galaxies with optical morphologies between irregulars and spheroidals with little sign of star formation. All three galaxies show some deviation from a smooth elliptical shape; VCC 190 is asymmetrical, VCC 611 has ongoing central star formation, and VCC 1533 is box-like in shape. Transition-type dwarfs may be between bursts of star formation; our sample additionally may have been weakly stripped ($<50$\% gas removal) while still in a group environment. Our sample is more gas rich and has a higher $M_*$ than is typical for dwarf transitionals (\citealt{Weisz2011a}; \citealt{DeLooze2013}), and so we could be observing the most extreme galaxies in the transition population.\\
\\
\textbf{Ambiguous evidence for gas accretion.} {In elliptical galaxies with accreted gas, recent or ongoing star formation is generally observed, and the \HI\ is in the form of clouds, streams, or a perturbed rotating disk. For two galaxies in our sample, we do observe star formation: the SDSS spectrum of VCC 611 shows strong centrally-located star formation, and the overall color of VCC 1533 is at the edge of what is expected for star-forming dwarfs (NUV$-r=3$). With single-dish spectra, we can only imperfectly probe the \HI\ morphology, but an \HI\ disk will have a clear two-horned profile, and both disks and streams have wide velocity widths even in dwarfs ($W_{50}\gtrsim50$ \kms). The \HI\ lines of VCC 611 and VCC 1533 galaxies are both narrow ($W_{50}<30$ \kms) and lack evidence of rotation as both are well-fit by a gaussian profile. VCC 190 is not star-forming, and the \HI\ at its optical position is similarly non-rotating. A tail is evident in \HI\ imaging of VCC 190, but is more consistent with gas removal than accretion.}\\
\\
\textbf{Evidence of tidal gas removal in VCC 190.} An observed \HI\ tail in VCC 190 points towards NGC 4224, while the gas coincident with the optical galaxy is practically undisturbed (fit well by a gaussian with $W_{50} = 15$ \kms). The tail is redshifted relative to the \HI\ in the optical galaxy, as would be expected from an encounter with NGC 4224, which is at a higher redshift than VCC 190. These all suggest that the tail is a tidal feature from an encounter approximately 1 Gy ago.
\\
\section*{Acknowledgements}
\noindent
We are extremely grateful for the observations performed by the faculty and students of the Undergraduate ALFALFA Team (UAT), without which this work would not be possible.  They performed the observations at Arecibo to confirm each galaxy's \HI, and performed the optical observations at KPNO of VCC 190 and VCC 611. The UAT is supported by NSF grant AST-1211005.\\
\\
ALFALFA has been supported by NSF grant AST-1107390, and grants from the Brinson Foundation.\\
\\
This work is based in part on observations made with the Karl G. Jansky Very Large Array, a facility of the National Radio Astronomy Observatory (NRAO). The NRAO is a facility of the National Science Foundation operated under cooperative agreement by Associated Universities, Inc.\\
\\
This work is based in part on observations made with the Arecibo Observatory. The Arecibo Observatory is operated by SRI International under a cooperative agreement with the National Science Foundation (AST-1100968), and in alliance with Ana G. M\'{e}ndez-Universidad Metropolitana, and the Universities Space Research Association.\\
\\
This work is based in part on observations made with the WIYN 0.9m telescope operated by WIYN
Inc. on behalf of a consortium of nine partner Universities and Organizations.

\end{document}